\documentstyle[11pt,epsf]{article}
 1
 1
 1

\def\lsim{\mathrel{\raise.3ex\hbox{$<$\kern-.75em\lower1ex\hbox{$\sim$}}}}
\def\gsim{\mathrel{\raise.3ex\hbox{$>$\kern-.75em\lower1ex\hbox{$\sim$}}}}
\newcommand{\as}{\alpha_s}
\begin{document}
\noindent
\thispagestyle{empty}
\renewcommand{\thefootnote}{\fnsymbol{footnote}}
\begin{flushright}
{\bf TTP96-35}\\
{\bf DTP/96/80}\\
{\bf hep-ph/9609411}\\
{\bf February 1997}\\
\end{flushright}
  \vspace{0.5cm}
\begin{center}
  \begin{Large}
Extracting $\alpha_s$ from Electron--Positron Annihilation 
around $10$ GeV
  \end{Large}
  \vspace{1.0cm}

\begin{Large}
 K.G.~Chetyrkin$^{a}$,
 J.H.~K\"uhn$^{b}$, 
 T.~Teubner$^{c}$
\end{Large}
\vskip -0.5 cm
\begin{itemize}
\begin{center}
\item[$^a$]
   Max-Planck-Institut f\"ur Physik, Werner-Heisenberg-Institut,\\
   F\"ohringer Ring 6, D-80805 Munich, Germany
\item[$^b$]
   Institut f\"ur Theoretische Teilchenphysik,\\ 
   Universit\"at Karlsruhe, 
   D-76128 Karlsruhe, Germany\\  
\item[$^c$]
   Department of Physics,\\
   University of Durham,   Durham, DH1 3LE, UK
\end{center}
\end{itemize}

  \vspace{0.7cm}
  {\bf Abstract}\\
\vspace{0.3cm}
\noindent
\begin{minipage}{15.0cm}
\begin{small}
The total cross section for electron--positron annihilation into
hadrons is calculated in the region just below the $B$ meson
threshold. QCD corrections up to third order, quark mass effects, 
initial state radiation, the 
running of $\alpha_{\rm QED}$ and the tails of the $\Upsilon$ resonances 
are included in the prediction. For given $\alpha_s$ the
prediction is accurate to 0.5 percent. An experimental measurement 
with the corresponding precision would allow to determine $\alpha_s$ 
with high accuracy at intermediate energies.
\end{small}
\end{minipage}
\end{center}
\vspace{1.2cm}
%
%
%
\noindent
{\Large 1. Introduction}\\

\noindent
One of the most precise and from the theoretical point unambiguous 
determinations of the strong coupling constant $\alpha_s$ has been
obtained from the hadronic decay rate of the $Z$ boson. Ingredients are
the large event rate which allows to reduce the statistical
uncertainty in $\alpha_s$ down to $0.003$, the precise calibration
through the accurate measurement of the luminosity or the leptonic
rate \cite{Blondel} and, last but not least, the elaborate theoretical
calculations \cite{YellowReport,CKKRep}. 
This high energy determination of $\alpha_s$
should and could be complemented by a similar measurement of $R_{\rm
had}$ in the energy region just below the $B$ meson threshold: large
event rates are available at the $e^+ e^-$ storage ring CESR and an
experimental analysis is currently underway \cite{DaveBessonpriv}.
Theoretical predictions for $R_{\rm had}$ below the $B{\overline B}$
threshold have been presented earlier \cite{rhad}. These include
terms up to order $\alpha_s^3$ for the massless limit and for $m_c^2/s$
corrections as well, terms of order $\alpha_s^2$ for the quartic mass
terms $(m_c^2/s)^2$, terms of order $\alpha_s (m_c^2/s)^3$ and
contributions from virtual bottom quarks up to order $\alpha_s^2 (s/m_b^2)$.
Recently it has been demonstrated that this expansion in $(m_c^2/s)$
provides an excellent approximation not only around $\sqrt{s} =
10$~GeV, but also down to 6 or even 5 GeV. 
\par
The forthcoming experimental analysis necessitates on the one hand a
thorough understanding of the background, e.g. from two photon events
or the feedthrough from $\tau$ pairs into the hadronic cross section,
and, on the other hand, a realistic prediction for the annihilation
channel. The former is outside the scope of this paper, the latter is
the subject of this work. It requires to
incorporate the effect of the running QED coupling constant and
initial state radiation. The photon vacuum polarization leads to an 
increase of the cross section by about 7--8\%. Magnitude and even sign 
of initial state corrections depend on the experimental procedure, 
in particular on the minimal mass of the hadronic system accepted 
for the event sample. Another
important issue is the treatment of the tails of the $\Upsilon$ 
resonances, which
contribute coherently through mixing with the photon and
incoherently through their radiative tails. In order to study these
effects and the associated theoretical uncertainties separately 
the corresponding analysis for the reaction $e^+ e^- \to \tau^+ \tau^-$ 
is presented in section 2. These results may also serve as an 
independent experimental
calibration of the cross section. The corresponding predictions for
the hadronic cross section are discussed in section 3. In section 4
we comment on the accuracy of the presented approach and section 5 
contains our summary and conclusions.
\vskip 1 cm
\noindent
{\Large 2. Lepton Pair Production}\\

\noindent
To calibrate the predictions for the hadron production cross section 
it seems appropriate and useful to calculate in a first step the cross 
section for lepton pair production. To arrive at a reliable 
prediction, initial and final state radiation, the effect of vacuum 
polarization and the influence of the nearby $\Upsilon$ resonances must 
be included. In this section results are presented for $\tau$ pair production.

\vskip 0.5 cm
\noindent
{\bf (a) Initial state radiation}

\noindent
The most important correction to the total cross section is introduced 
by initial state radiation. It leads to a reduction of the invariant 
mass of the produced lepton pair or hadronic system and, for fairly 
loose cuts, to a significant enhancement of the cross section, albeit 
with events of significantly lower invariant mass of the system of
interest. For the precise determination of $R_{\rm had}$ discussed below 
it is advisable to exclude the bulk of these low mass events. This reduces the 
size of the correction and at the same time the dependence on the 
input for $R_{\rm had}$ from the lower energy region. 

The treatment of initial state radiation has advanced significantly as 
a consequence of the detailed calculations performed for the analysis
of the $Z$ line shape. The result for the inclusive cross section can 
be written in the form 
\begin{equation}
\sigma(s) = \int_{z_0}^1 {\rm d}z\,\sigma_0(sz)\,G(z)\,.
\label{eqisrint}
\end{equation}
The cross section including photon vacuum polarization (running $\alpha$) 
is denoted by $\sigma_0$. The invariant mass of the produced fermion 
pair is given by $s\,z$, where 
\begin{equation}
\frac{m_{\rm min}^2}{s} \leq z_0 \leq z \leq 1
\label{eqisrborder}
\end{equation}
and $m_{\rm min} = 2 m_\ell$ for lepton pairs and $m_{\rm min} = 2 m_\pi$ 
for hadron production. In the cases of interest for this paper $z_0$ 
will have to be adopted to the experimental setup. Typically it is 
significantly larger than the theoretically allowed minimal value. 
For $E_{\rm cm} = 10.52$~GeV a cut in $z$ around $0.25$ corresponding 
to roughly $5$~GeV in the minimal mass will in the case of hadronic 
final states exclude the charmonium resonance region and the charm 
threshold region as well, thus limiting the hadron analysis to truely 
multihadronic final states. 

The complete radiator function $G(z)$ up to order $\alpha^2$ has been 
calculated in \cite{Burgers,vanNeerven}. The resummation of leading 
logarithms is discussed in \cite{YRBerends} (see also \cite{Kuraev}). 
For the present purpose an approximation is adequate which is exact in 
order $\alpha$ and which includes the dominant terms of order 
$\alpha^2$ plus leading logarithms. For the radiator function $G$ we 
thus take \cite{JadachWard}
\begin{equation}
G(z) = \beta (1-z)^{\beta-1}\,{\rm e}^{\delta_{yfs}}\,F\,\left(
 \delta_C^{V+S} + \delta_C^H \right)\,,
\label{eqGc}
\end{equation}
with
\begin{eqnarray}
\beta &=& \frac{2\alpha}{\pi}(L-1)\,,\nonumber\\
L &=& \ln\frac{s}{m_e^2}\,,\nonumber\\
\delta_{yfs} &=& \frac{\alpha}{\pi} \left( \frac{L}{2} - 1 + 2\zeta(2) 
 \right)\,,\nonumber\\
\delta_c^{V+S} &=& 1+\frac{\alpha}{\pi}(L-1)+\frac{1}{2}
 \left(\frac{\alpha}{\pi}\right)^2 L^2\,,\nonumber\\
\delta_C^H &=& -\frac{1-z^2}{2}+\frac{\alpha}{\pi}L\left[-\frac{1}{4}
 \left(1+3z^2\right)\ln z -1+z\right]\,,\nonumber\\
F &=& \frac{{\rm e}^{-\beta\gamma_E}}{\Gamma(1+\beta)}\,,\nonumber
\end{eqnarray}
where $\gamma_E = 0.5772\ldots$ is Euler's constant. 
Eq.~(\ref{eqGc}) is suited for quick numerical integration. 
The difference to the initial state radiation convolution using 
the complete order $\alpha^2$ result (eq~(3.12) of \cite{YRBerends}) 
is below one permille. 
The predictions for the reference energy of $10.52$~GeV and a 
variety of cuts $z_0$ are listed in Table~\ref{table1}. 
\begin{table}[htb]
\caption{Dependence of the cross section $e^+ e^- \to \tau^+ \tau^-$ 
on the cutoff $z_0$ for $E_1 = 10.52$~GeV.}
\begin{center}
\begin{tabular}{|c||c|c|c|c|c|c|c|} \hline
$z_0$ 
 & 0.114       & 0.20   & 0.25   & 0.5    & 0.8    & 0.9    & 0.95 
\\ \hline
$m_{\rm cut}\ [GeV]$ 
 & $2m_{\tau}$ & 4.705  & 5.260  & 7.439  & 9.409  & 9.980  & 10.254 
\\ \hline\hline
$\sigma\ [nb]$ 
 & 0.9269      & 0.9101 & 0.8998 & 0.8543 & 0.7831 & 0.7359 & 0.6922 
\\ \hline
$\sigma/\sigma_{pt}$ 
 & 1.1822      & 1.1608 & 1.1477 & 1.0896 & 0.9988 & 0.9386 & 0.8829 
\\ \hline
\end{tabular}
\end{center}
\label{table1}
\end{table}
\begin{figure}[htb]
\begin{center}
\leavevmode
\epsfxsize=17.cm
\epsffile[70 290 510 530]{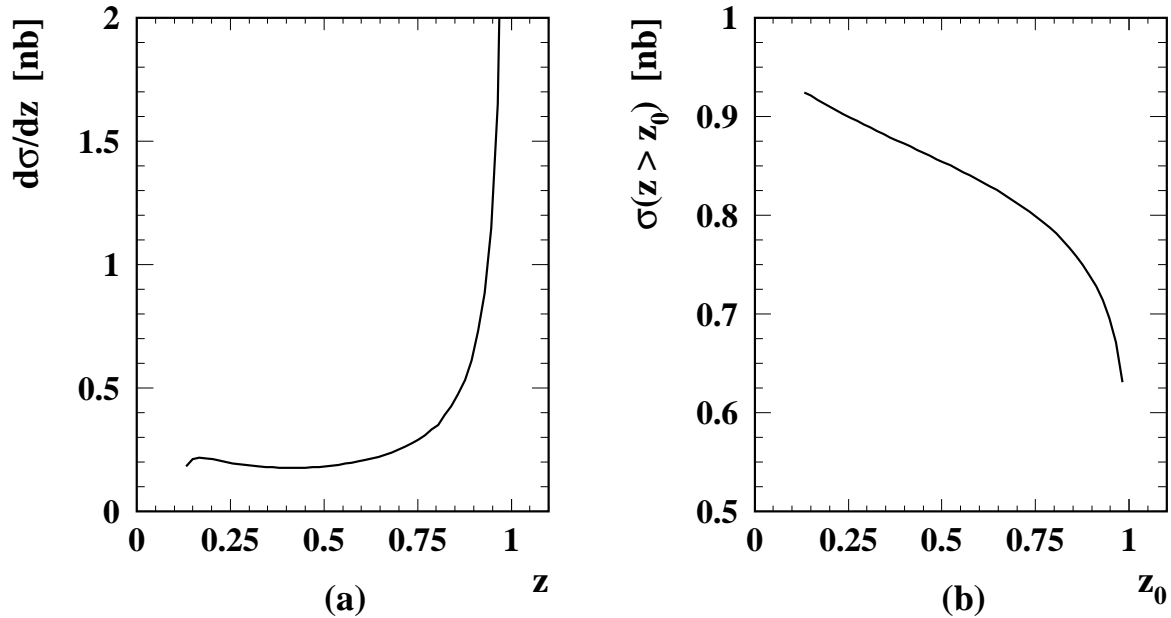}
\vskip -3mm
\caption[]{\label{fig1ab} {\em (a) The differential cross section 
${\rm d}\sigma(e^+ e^- \to \tau^+ \tau^-)/{\rm d}z$ and (b) the 
cutoff dependent cross section $\sigma(z > z_0)$ in nb for 
the centre of mass energy $E_1 = 10.52$~GeV as functions of $z$ and 
$z_0$, respectively.}} 
\end{center}
\end{figure}
The differential distribution ${\rm d}\sigma/{\rm d}z$ and the 
cutoff dependent cross section $\sigma(z > z_0)$ are displayed 
in Fig.~\ref{fig1ab}a, b, respectively, as functions of
$z$, $z_0$. For the cross section before convolution the corrections from 
the vacuum polarization, the tails of the $\Upsilon$ resonances and 
final state radiation are included. 
Initial state radiation of lepton pairs and of hadrons has been 
treated in \cite{vanNeerven,KKKS}. These could easily be incorporated 
in the present formalism. The resulting corrections are at the 
permille level and will be ignored for the moment.
Corrections from initial state radiation are evidently huge for loose
or extremely tight cutoffs, with opposite sign. The strong cutoff
dependence of the result is apparent from Table \ref{table1} and
Fig.~\ref{fig1ab} as well. The result is relatively stable against
variations of $z_0$ for values of $z_0$ below 0.75. At the
same time the sensitivity to the input for $\sigma_0$ from smaller 
$s' = s\,z$ is eliminated for this choice. 
From this viewpoint a value around 0.7 to 0.8 seems optimal for
the experimental determination of $R_{\rm had}$. On the other
hand, if one allows to use input from smaller $s'$ from other experiments
or from the measured events with photons from initial state radiation,
then also lower values of $z_0$ down to about 0.25 are equally
acceptable. 

The distribution ${\rm d}\sigma/{\rm d}z$ is flat in the 
region $0.25 < z < 0.75$ and numerically small. An experimental
analysis based on a cutoff $z_0 = s'/s$ with a reasonably symmetric
resolution in $s'$ is thus insensitive towards the details of the
resolution function. In fact, even an uncontrolled shift of the
central value by $\Delta z_0 = 0.05$ would lead to a 1\% deviation 
of the cross section only.

\noindent
{\bf (b) Final state radiation}

\noindent
Initial state radiation is strongly enhanced, a consequence of the 
inherently large logarithm in the correction 
function (cf. eq.~(\ref{eqGc})). Final state radiation with loose 
cuts, in 
contrast, is typically of order $\alpha/\pi$ without a large logarithm. 
In the massless limit and without a cutoff on the invariant fermion 
pair mass the correction factor $(1 + 3/4\,\alpha/\pi)$ amounts 
to two permille only. It depends, however, on the lepton mass 
and the cutoff. For the relatively loose cuts with $z_0$ between 
0.25 and 0.8 one may well use the totally inclusive correction 
function $r^{(1)}$ defined by 
\begin{equation}
R_{f\bar f}\, := \, \frac{\sigma(e^+e^-\to\gamma^*\to f\bar f\ldots)}
                      {\sigma_{pt}}
 \quad = \quad r^{(0)} + 
               \,\bigg(\frac{\alpha}{\pi}\bigg)\,r^{(1)} + \ldots\,,
\label{eqrdef}
\end{equation}
with $\ \sigma_{pt} = 4\pi\alpha^2/3s\ $
and
\begin{eqnarray}
r^{(0)} & = & \frac{\beta}{2}\,(3-\beta^2)\,,\\[2mm]
r^{(1)} & = &
\frac{\left( 3 - {\beta^2} \right) \,\left( 1 + {\beta^2} \right) }{2
    }\,\bigg[\, 2\,\mbox{Li}_2(p) + \mbox{Li}_2({p^2}) + 
     \ln p\,\Big( 2\,\ln(1 - p) + \ln(1 + p) \Big) 
      \,\bigg] \,\nonumber\,\\ 
 & & \mbox{} - 
  \beta\,( 3 - {\beta^2} ) \,
   \Big( 2\,\ln(1 - p) + \ln(1 + p) \Big)  - 
  \frac{\left( 1 - \beta \right) \,
     \left( 33 - 39\,\beta - 17\,{\beta^2} + 7\,{\beta^3} \right) }{16}\,
   \ln p\,\nonumber\,\\ 
 & & \mbox{} + 
  \frac{3\,\beta\,\left( 5 - 3\,{\beta^2} \right) }{8}
\,,
\end{eqnarray}
where
\begin{eqnarray}
  p \, = \, \frac{1-\beta}{1+\beta}\,,\qquad\,
    \beta \, = \, \sqrt{1-4m_f^2/s}\,.
\end{eqnarray}
In the case of hadron production, the main motivation 
of this investigation, photons from final state radiation will 
in general anyhow be included in the hadronic invariant mass. 

\vskip 0.5 cm
\noindent
{\bf (c) Leptonic and hadronic vacuum polarization}

\noindent
The leptonic vacuum polarization in one loop approximation is given 
by 
\begin{equation}
\widehat\Pi_{\gamma\gamma}(s) := 
\Pi_{\gamma\gamma}(s) - \Pi_{\gamma\gamma}(0) =  
\frac{\alpha}{3\pi}\,\sum_f N_c \, Q_f^2 \, P(s, m_f)\,,
\label{eqvacpollepdef}
\end{equation}
with
\begin{equation}
P(s, m_f) = \frac{1}{3} - \left(1+\frac{2m_f^2}{s}\right)\,
                          \left(2+\beta\ln\frac{\beta-1}{\beta+1}\right)\,.
\label{eqvacpollepp}
\end{equation}
For $|s| \gg m_{\ell}^2$ the function $P(s, m_f)$ is well approximated by
\begin{equation}
P_{\rm asymp.}(s, m_f) = -\frac{5}{3} + \ln\bigg(-\frac{s}{m_f^2}
 + i \,\epsilon\bigg)\,.
\label{eqvacpolleppappr}
\end{equation}
The hadronic vacuum polarization is obtained from it's imaginary part 
via a dispersion relation 
\begin{equation}
{\rm Re}\widehat\Pi_{\rm had}(q^2) = \frac{\alpha q^2}{3\pi} 
\, {\rm P} \int_{m_{\pi}^2}^{\infty} \frac{R_{\rm had}(s')}{s'(s'-q^2)}
\, {\rm d}s'\,.
\label{eqvacpolhad}
\end{equation}
To avoid complications that arise from the numerical evaluation of 
the dispersion integral over the data in the timelike region, the 
integral is evaluated at the corresponding value of $s$ in the 
spacelike region. 
We use parametrisations from the evaluation of eq.~(\ref{eqvacpolhad}) 
in the spacelike region as provided by \cite{jegerl,burkhardt}. 
The results for the 
complete cross sections calculated with the different parametrisations 
from \cite{jegerl} and \cite{burkhardt} agree to better 
than $10^{-4}$ in the region of interest. The error from the 
identification of spacelike and 
timelike $\widehat\Pi_{\rm had}(q^2)$ should be 
small except for 
the $b$ quark contribution, where the threshold is fairly close to the
$q^2$ values of interest. Therefore we subtract the perturbative
$b$ quark contribution evaluated for spacelike $q^2$ and add the
corresponding value for timelike $q^2 = s$. We have checked that this
ansatz is in excellent agreement with the numerical
evaluation\footnote{We thank H.~Burkhardt for providing the numerical
evaluation of (\ref{eqvacpolhad}) required for this comparison.} of
(\ref{eqvacpolhad}) in the timelike region. 
Effects due to $\Upsilon$ resonances are described in detail below. 

\begin{figure}[htb]
\begin{center}
\leavevmode
\epsfxsize=12.cm
\epsffile[110 280 460 560]{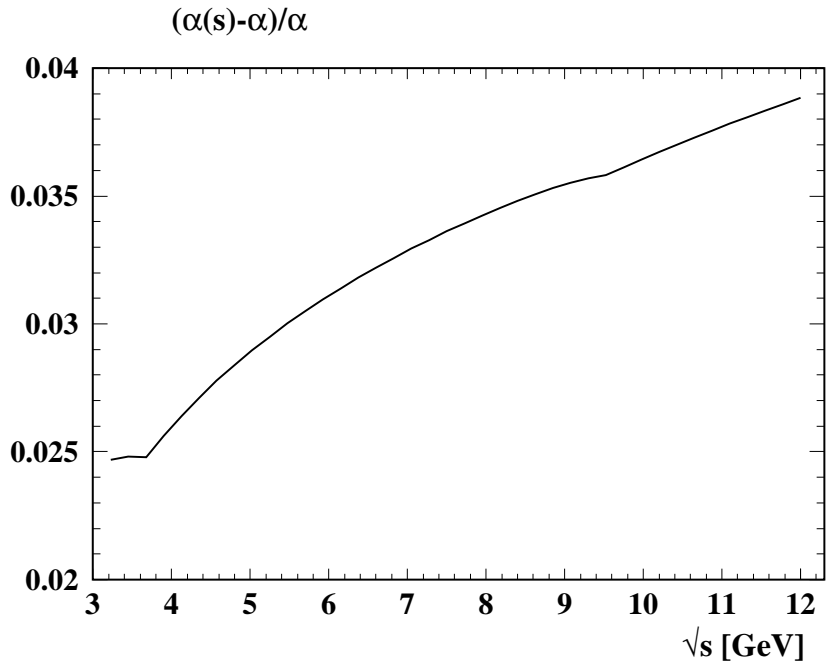}
\vskip -3mm
\caption[]{\label{figalpha} {\em The relative shift in $\alpha(s)$ as 
a function of $\sqrt{s}$ from the hadronic plus leptonic vacuum 
polarization as described in the text.}} 
\end{center}
\end{figure}

The running $\alpha$ is then obtained from the real part of $\widehat\Pi$ 
through 
\begin{equation}
\alpha(s) = \frac{\alpha}{1-{\rm Re}\widehat\Pi(s)}\,.
\label{eqalpharunning}
\end{equation}
The relative shift in $\alpha(s)$ from the 
hadronic plus leptonic vacuum polarization is shown in 
Fig.~\ref{figalpha} as a function of $\sqrt{s}$. The size 
of the individual contributions is 
listed in Table~\ref{table2} for the reference energy 
$E_1=10.52$ GeV and for a few selected lower energies. 
The uncertainty in the cross section from our treatment of the
hadronic vacuum polarization in the timelike region is estimated 
to be below two permille.
\begin{table}[htb]
\caption{Individual contributions to ${\rm Re}\widehat\Pi(s) \cdot 10^2$.}
\begin{center}
\begin{tabular}{|c||c|c|c|c|c|} \hline
$\sqrt{s}/$GeV & $2m_{\tau}$ & 5      & 7      & 9      & 10.52  \\ \hline
$e$            & 1.241       & 1.294  & 1.346  & 1.385  & 1.409  \\
$\mu$          & 0.415       & 0.468  & 0.520  & 0.559  & 0.583  \\
$\tau$         & -0.207      & -0.049 & 0.047  & 0.101  & 0.132  \\
had            & 0.923       & 1.096  & 1.269  & 1.379  & 1.456  \\ \hline
$1/\alpha(s)$  & 133.79      & 133.19 & 132.68 & 132.34 & 132.13 \\ \hline
\end{tabular}
\end{center}
\label{table2}
\end{table}

\vskip 0.5 cm
\noindent
{\bf (d) Narrow $\Upsilon$ resonances}

\begin{table}[htb]
\caption{Relative contributions from the $\Upsilon$ resonances to 
the leptonic cross section $\sigma(e^+ e^- \to \tau^+ \tau^-)$ 
for the energies $E_1 = 10.52$~GeV and $E_2 = 9.98$~GeV and two
different values of the cutoff $m_{\rm min}$. Interference
terms and radiative tails are listed separately. Also given are the
continuum contributions and the resulting predictions for the total
cross section.}
\begin{center}
\begin{tabular}{|c|c|c|c|c|c|c|} \hline
                  &$\Upsilon(1)$&$\Upsilon(2)$&$\Upsilon(3)$&
                   $\Upsilon(4)$&$\Upsilon(5)$&$\Upsilon(6)$\\ \hline\hline
$M\ [GeV]$ & 
 $9.460$  & $10.023$ & $10.355$ & $10.580$ & $10.865$ & $11.019$ \\ \hline
$\Gamma_e\ [keV]$ &
 $1.32$   & $0.576$  & $0.476$  & $0.24$   & $0.31$   & $0.13$ \\ \hline
$\Gamma_{\rm tot}\ [MeV]$ &
 $0.0525$ & $0.044$  & $0.0263$ & $23.8$   & $110$    & $79$ \\ \hline\hline
 \multicolumn{7}{|l|}{$E_1 = 10.52$ GeV, $m_{\rm min}=2m_{\tau}=3.554$ GeV:} 
 \\ \hline
Continuum:  & \multicolumn{6}{c|}{$0.9256$ nb} \\ \hline
Interf.: 
 & $2.7\cdot 10^{-4}$   & $2.9\cdot 10^{-4}$  & $7.5\cdot 10^{-4}$ 
 & $-1.04\cdot 10^{-3}$ & $-3.0\cdot 10^{-4}$ & $-9.7\cdot 10^{-5}$ \\ \hline
$\sum$ interf.:    & \multicolumn{6}{c|}{$-1.3\cdot 10^{-4}$} \\ \hline
Rad.~tails~: 
 & $5.5\cdot 10^{-4}$ & $2.5\cdot 10^{-4}$ & $8.1\cdot 10^{-4}$ & 
   --                 & --                 & --                     \\ \hline
$\sum$ rad. tails: & \multicolumn{6}{c|}{$1.61 \cdot 10^{-3}$} \\ \hline
Total: & \multicolumn{6}{c|}{$0.9269$ nb} \\ \hline\hline
 \multicolumn{7}{|l|}{$E_1 = 10.52$ GeV, $m_{\rm min}=5$ GeV:} \\ \hline
Continuum:  & \multicolumn{6}{c|}{$0.9033$ nb} \\ \hline
Interf.: 
 & $2.9\cdot 10^{-4}$   & $3.1\cdot 10^{-4}$  & $7.7\cdot 10^{-4}$ 
 & $-1.07\cdot 10^{-3}$ & $-3.1\cdot 10^{-4}$ & $-9.8\cdot 10^{-5}$ \\ \hline
$\sum$ interf.:    & \multicolumn{6}{c|}{$-9.7\cdot 10^{-5}$} \\ \hline
Rad.~tails~: 
 & $5.7\cdot 10^{-4}$ & $2.6\cdot 10^{-4}$ & $8.3\cdot 10^{-4}$ & 
   --                 & --                 & --                     \\ \hline
$\sum$ rad. tails: & \multicolumn{6}{c|}{$1.65 \cdot 10^{-3}$} \\ \hline
Total: & \multicolumn{6}{c|}{$0.9047$ nb} \\ \hline\hline
 \multicolumn{7}{|l|}{$E_2 = 9.98$ GeV, $m_{\rm min}=2m_{\tau}=3.554$ GeV:} 
 \\ \hline
Continuum:  & \multicolumn{6}{c|}{$1.0200$ nb} \\ \hline
Interf.: 
 & $6.4\cdot 10^{-4}$ & $-3.51\cdot 10^{-3}$ & $-4.5\cdot 10^{-4}$ & 
   $-1.6\cdot 10^{-4}$ & $-1.6\cdot 10^{-4}$ & $-5.9\cdot 10^{-5}$ \\ \hline
$\sum$ interf.:    & \multicolumn{6}{c|}{$-3.70\cdot 10^{-3}$} \\ \hline
Rad.~tails: 
 & $1.08\cdot 10^{-3}$ & -- & -- &  -- & -- & -- \\ \hline
$\sum$ rad. tails: & \multicolumn{6}{c|}{$1.08\cdot 10^{-3}$} \\ \hline
Total: & \multicolumn{6}{c|}{$1.0173$ nb} \\ \hline\hline
 \multicolumn{7}{|l|}{$E_2 = 9.98$ GeV, $m_{\rm min}=5$ GeV:} \\ \hline
Continuum:  & \multicolumn{6}{c|}{$0.9947$ nb} \\ \hline
Interf.: 
 & $6.8\cdot 10^{-4}$ & $-3.60\cdot 10^{-3}$ & $-4.6\cdot 10^{-4}$ & 
   $-1.6\cdot 10^{-4}$ & $-1.6\cdot 10^{-4}$ & $-5.8\cdot 10^{-5}$ \\ \hline
$\sum$ interf.:    & \multicolumn{6}{c|}{$-3.75\cdot 10^{-3}$} \\ \hline
Rad.~tails: 
 & $1.11\cdot 10^{-3}$ & -- & -- &  -- & -- & -- \\ \hline
$\sum$ rad. tails: & \multicolumn{6}{c|}{$1.11\cdot 10^{-3}$} \\ \hline
Total: & \multicolumn{6}{c|}{$0.9920$ nb} \\ \hline
\end{tabular}
\end{center}
\label{table3}
\end{table}
\noindent
Up to this point we have not included the contributions from the 
$\Upsilon$ resonances. As a consequence of their close proximity 
their effect will be enhanced and thus should 
be discussed separately. The Breit--Wigner amplitudes from the 
$\Upsilon$ resonances above and below the $B{\overline B}$ threshold 
interfere with the virtual photon amplitude and thus enhance or 
decrease the cross section by
\begin{equation}
\delta\sigma_{\rm int}= 
\frac{6\,\Gamma_e}{\alpha(s)\,s} \, 
\frac{M^3(s-M^2)}{(s-M^2)^2+\Gamma^2 M^2}\,\sigma_0\,.
\label{eqbwint}
\end{equation}
The individual contributions are listed in Table~\ref{table3} for the 
two cms energies $E_1 = 10.52$~GeV and $E_2 = 9.98$~GeV after the
convolution with initial state radiation. For $E_1$ these interference 
terms happen to cancel to a large extent, leaving a minute correction 
with a relative magnitude around $10^{-4}$. For $E_2$, however, due to 
the close proximity of $\Upsilon(2S)$ one receives a non--negligible 
negative contribution of about $-0.4$\%. This effect could become 
relevant in precision tests. 
\par
Also the radiative tails of the resonances with $M < \sqrt{s}$ have to 
be taken into consideration. These can be easily calculated with the
radiator function eq.~(\ref{eqGc}), taking either the full
Breit--Wigner resonance 
\begin{equation}
\delta\sigma_{\rm R} = 
\left(\frac{3\,\Gamma_e\,M}{\alpha(s)\,s}\right)^2  
\frac{M^4}{(s-M^2)^2+\Gamma^2 M^2}\,\sigma_0
\label{eqbwtailfull}
\end{equation} 
or the narrow width approximation
\begin{equation}
\delta\sigma_{\rm R} \Big|_{\rm NW} = 
\frac{9\,\Gamma_e^2}{\alpha(s)^2} \frac{M}{\Gamma} \,\pi\, \delta(s-M^2)
\,\sigma_0\,. 
\label{eqbwtailnarroww}
\end{equation} 
Both formulae lead to nearly identical results. The individual 
contributions are also listed in Table~\ref{table3}. 
In total they raise the lepton pair cross section by about 0.16\% or 0.11\%
for $E_1$ and $E_2$, respectively. 
At this point a brief comment ought to be made concerning the 
treatment of $\Upsilon(4S)$. The approximation of neglecting the 
energy dependence of the width of the resonance 
is justified even for $E_1$, since the nominal width $\Gamma = 24$~MeV is 
significantly smaller than the difference between the cms energy $E_1$ 
and the mass of the resonance: $\Delta E = 60$~MeV. 
\pagebreak

\noindent
{\Large 3. The Total Cross Section for Hadron Production}\\

\noindent
As stated already in the introduction the emphasis of this work is on 
the energy just below the bottom meson threshold. Virtual $b$ quark 
loops can be easily taken into account, and in fact it has been 
demonstrated that the hard mass expansion works surprisingly well 
for these diagrams --- not only far below but even close to the 
nominal $b$ quark threshold. On the other hand this energy is 
sufficiently far above the open charm threshold such that charm 
quark mass effects can be included through an expansion in powers 
of $m_c$, if quadratic and quartic terms are incorporated. With 
this method the production cross section can be predicted reliably 
not only in the high energy region but even relatively close to 
threshold through an expansion in $m^2/s$. 
This approach was suggested originally in \cite{quart,rhad}. The 
calculation of $R_{\rm had}$ to second order, including the full mass 
dependence \cite{hkt1} has demonstrated the nearly perfect agreement 
between approximate and exact result for the coefficient of the 
$\alpha_s^2$ term for $E_{\rm cm}$ above $4m$. 
In fact, even for $E_{\rm cm}$ around $3m$, which is around the lowest
advisable value of the cutoff $m_{\rm min}=5$ GeV, the deviation of the 
$\alpha_s^2$ coefficient from the complete result leads to a 
difference of less than five percent in the rate. For the energy 
around 10 GeV this region contributes through the radiative tail 
only, and the approximation is thus adequate throughout. In total 
we thus use the following individual contributions
\begin{equation}
R = R_{\rm NS} + R_{\rm S} + \delta R_{m_b} + \delta R_{m_c} + 
 \delta_{\rm QED}\,,
\label{eqRhad}
\end{equation}
where
\begin{eqnarray}
R_{\rm NS} & = & \sum_{f = u,d,s,c}  3 \, Q_f^2 
\left[
1 +  \frac{\as}{\pi}  
+ 1.5245
\left(\frac{\as}{\pi}\right)^2
-11.52033
\left(\frac{\as}{\pi}\right)^3\,
\right]
\,,\nonumber\\
R_{\rm S} & = &
-\left(\frac{\as}{\pi}\right)^3
\Big(\sum_{u,d,s,c} Q_f\Big)^2 \, 1.239 \ = \ 
-0.55091
\left(\frac{\as}{\pi}\right)^3
\,,\nonumber\\
\delta R_{m_b} & = & \sum_{f = u,d,s,c}  3 \, Q_f^2     
\left(\frac{\as}{\pi}\right)^2 \frac{s}{{\overline m}_b^2}
\left[
\frac{44}{675}   +  \frac{2}{135} \log \frac{{\overline m}_b^2}{s}
\right] 
\,,\nonumber\\
\delta R_{m_c} & = &
3\,Q_c^2 \,12\,\frac{m_c^2}{s} \frac{\as}{\pi}
\left[
1
+
9.097 
\frac{\as}{\pi}
+
53.453
\left(\frac{\as}{\pi}\right)^2
\right]
-
3 \sum_{f=u,d,s,c}Q_f^2\frac{m_c^2}{s}
 \left(\frac{\as}{\pi}\right)^3
   6.476  
\nonumber\\
& & +
3\,Q_c^2 \frac{m_c^4}{s^2}  
\left[
-6
-22
\frac{\as}{\pi}
+
\left(
141.329 - \frac{25}{6}\ln\frac{m_c^2}{s}
\right)
\left(\frac{\as}{\pi}\right)^2
\right]
\nonumber\\
& &+3 \sum_{f=u,d,s,c} Q_f^2
\frac{m_c^4}{s^2}  
\left(\frac{\as}{\pi}\right)^2
\left[
-0.4749
- \ln\frac{m_c^2}{s}
\right] 
-3\,Q_c^2 \frac{m_c^6}{s^3}  
\left[
8
+\frac{16}{27}
\frac{\as}{\pi}
\left(
6\ln\frac{m_c^2}{s} + 155
\right)
\right]\,,
\nonumber\\
\delta_{\rm QED} & = & 
\sum_{f=u,d,s,c} 3\,Q_f^4 \frac{\alpha}{\pi} \frac{3}{4}
\,.
\nonumber
\end{eqnarray}
The formulae are evaluated for $n_f=4$ with $\alpha_s$ and the charm 
quark mass interpreted accordingly. For the massless case and 
the $m^2/s$ terms the results are available in third order, for 
the quartic and $m^6/s^3$ terms in second and first order 
$\alpha_s$, respectively. Tables which list the numerical values of
the running quark masses and the magnitude of the individual 
contributions can be found in \cite{rhad,CKKRep} together with the 
details of the matching between the theories with $n_F = 4$ and $5$.

\begin{figure}[htb]
\begin{center}
\leavevmode
\epsfxsize=17.cm
\epsffile[60 280 520 540]{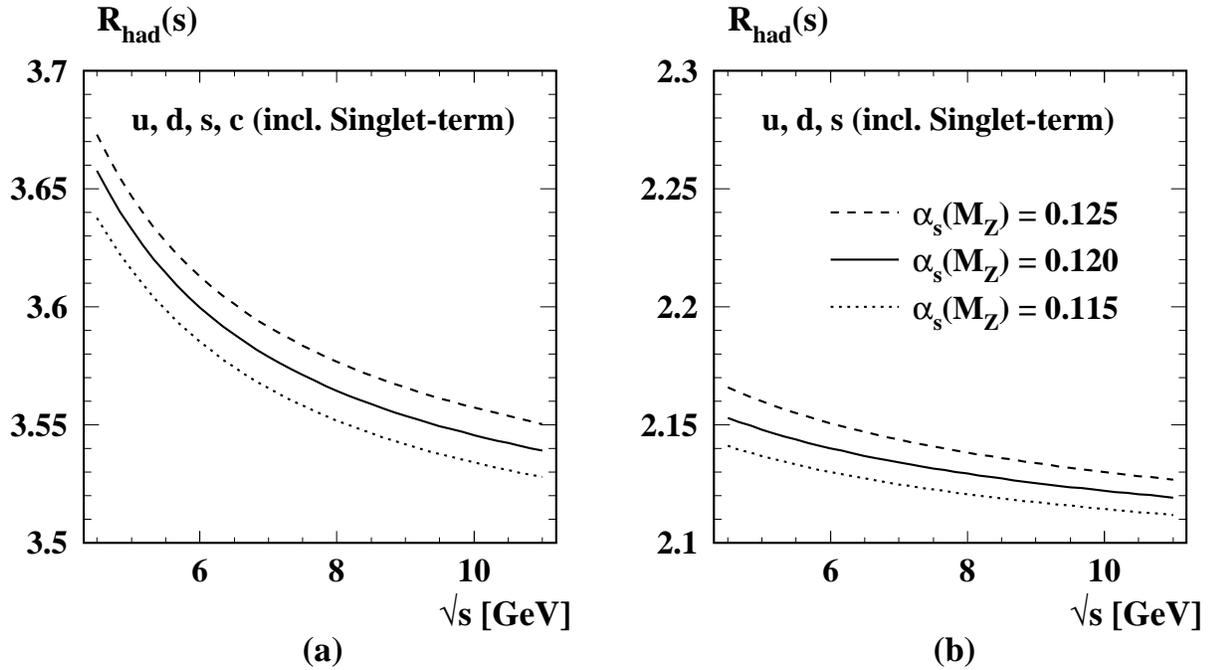}
\vskip -3mm
\caption[]{\label{fig3ab} {\em (a) $R_{\rm had}(s)$ as defined in 
eq.~(\ref{eqRhad}), and (b) only the contributions from the light 
($u$, $d$ and $s$) quark currents for different values of $\alpha_s$.}} 
\end{center}
\end{figure}
The predictions for $R_{\rm had}$ as a function of $E_{\rm cm}$ are shown 
in Fig.~\ref{fig3ab} for three values of the strong coupling constant,
$\alpha_s(M_Z^2) = 0.115$, $0.120$ and $0.125$. 
Fig.~\ref{fig3ab}a displays the contributions induced by the light plus 
charm quark currents (as defined in eq.~(\ref{eqRhad})), 
whereas in Fig.~\ref{fig3ab}b only the light ($u$, $d$ and $s$) quark 
current contributions are shown. Note, that this figure by definition 
does not contain 
QED--corrections from initial state radiation and vacuum polarization, 
but includes the tiny singlet terms, which cannot be attributed to one 
individual quark species and the final state QED--corrections 
$\delta_{\rm QED}$. 
The quark masses are chosen to be 
${\overline m}_c({\overline m}_c) = 1.24$~GeV ($n_f=4$) 
and ${\overline m}_b({\overline m}_b) = 4.1$~GeV ($n_f=5$) 
corresponding to pole masses of $1.6$~GeV and  $4.7$~GeV, 
respectively, if $\alpha_s^{(n_f=5)}(M_Z)=0.120$ is adopted. 
A variation of the charm quark mass around the default value 
by $+300$~MeV/$-300$~MeV changes 
$R_{\rm had}$ only by $+0.006/-0.004$ for $\sqrt{s} = 10.52$ GeV and by 
$+0.031/-0.028$ for $\sqrt{s} = 5$ GeV. 

Throughout this paper the QCD results are interpreted in fixed order
$\alpha_s^3$ without any attempt to improve the formulae through the
inclusion of guesses for higher order coefficients. An estimate of the
scale dependence is easily obtained through the evaluation of a
variant of eq.~(\ref{eqRhad}) where $R_{\rm NS}$ is calculated for a general
t'Hooft scale $\mu^2$. Adopting $\alpha_s(10.5\ {\rm GeV}) = 0.177$
corresponding to $\alpha_s(M_Z) = 0.12$ and varying $\mu^2$ between 
$s/4$ and $4 s$ the predicted value of $R$ varies by -2 and +0.2
permille. This is well below the anticipated experimental precision. 
Alternatively we may include in eq.~(\ref{eqRhad}) an $\alpha_s^4$
term with the coefficient based on a recent estimate 
in \cite{Kataev}. This would lead to a decrease in $R$ by 1.7 
permille, again far below the forseeable experimental accuracy.

Let us now discuss the impact of initial state radiation, the running
$\alpha$ and the $\Upsilon$ resonances on the hadronic cross section,
as observed at around $10$~GeV under realistic experimental
conditions. Lower energies contribute again through initial state
radiation (eq.~(\ref{eqisrint})). If we restrict the cutoff $z_0$ to a
value of $0.25$ which corresponds to a cut on the mass of the hadronic
system of around $5$~GeV, eq.~(\ref{eqRhad}) for $R_{\rm had}$ can be
applied also for $\sigma_0(z s')$ which appears in the integrand of 
(\ref{eqisrint}). At the same time this cutoff excludes the region of
the broad charmonium resonances which have not been well explored up
today. A cutoff around $z_0 = 0.7$ reduces the initial state radiation
corrections further and eliminates contributions from
the lower energy range completely. 
The vacuum polarization of the virtual photon has been discussed
before for the case of the $\tau$ lepton production and is identical
for the hadronic cross section. 
\begin{table}[htb]
\caption{Relative contributions from the $\Upsilon$ resonances to 
the hadronic cross section $\sigma(e^+ e^- \to {\rm hadrons})$ 
for the energies $E_1 = 10.52$~GeV and $E_2 = 9.98$~GeV and 
the two different values of the cutoff $m_{\rm min} = 5$~GeV and 9~GeV. 
Interference terms and radiative tails are listed separately. 
Also given are the continuum contributions and the resulting 
predictions for the total cross section.}
\begin{center}
\begin{tabular}{|c|c|c|c|c|c|c|} \hline
                  &$\Upsilon(1)$&$\Upsilon(2)$&$\Upsilon(3)$&
                   $\Upsilon(4)$&$\Upsilon(5)$&$\Upsilon(6)$\\ \hline\hline
 \multicolumn{7}{|l|}{$E_1 = 10.52$ GeV, $m_{\rm min}=5$ GeV:} \\ \hline
Continuum:  & \multicolumn{6}{c|}{$3.2228$ nb} \\ \hline
Interf.: 
 & $2.9\cdot 10^{-4}$ & $3.1\cdot 10^{-4}$ & $7.7\cdot 10^{-4}$ & 
   $-1.06\cdot 10^{-3}$ & $-3.1\cdot 10^{-4}$ & $-9.7\cdot 10^{-5}$ \\ \hline
$\sum$ interf.:    & \multicolumn{6}{c|}{$-1.0\cdot 10^{-4}$} \\ \hline
Rad.~tails: 
 & $5.88\cdot 10^{-3}$ & $5.38\cdot 10^{-3}$ & $1.212\cdot 10^{-2}$ & 
   -- & -- & -- \\ \hline
$\sum$ rad. tails: & \multicolumn{6}{c|}{$2.343\cdot 10^{-2}$} \\ \hline
Total: & \multicolumn{6}{c|}{$3.2980$ nb} \\ \hline\hline
 \multicolumn{7}{|l|}{$E_1 = 10.52$ GeV, $m_{\rm min}=9$ GeV:} \\ \hline
Continuum:  & \multicolumn{6}{c|}{$2.8515$ nb} \\ \hline
Interf.: 
 & $4.0\cdot 10^{-4}$ & $3.7\cdot 10^{-4}$ & $8.9\cdot 10^{-4}$ & 
   $-1.19\cdot 10^{-3}$ & $-3.3\cdot 10^{-4}$ & $-1.1\cdot 10^{-4}$ \\ \hline
$\sum$ interf.:    & \multicolumn{6}{c|}{$3.9\cdot 10^{-5}$} \\ \hline
Rad.~tails: 
 & $6.64\cdot 10^{-3}$ & $6.08\cdot 10^{-3}$ & $1.370\cdot 10^{-2}$ & 
   $5\cdot 10^{-5}$ & -- & -- \\ \hline
$\sum$ rad. tails: & \multicolumn{6}{c|}{$2.648\cdot 10^{-2}$} \\ \hline
Total: & \multicolumn{6}{c|}{$2.9272$ nb} \\ \hline\hline
 \multicolumn{7}{|l|}{$E_2 = 9.98$ GeV, $m_{\rm min}=5$ GeV:} \\ \hline
Continuum:  & \multicolumn{6}{c|}{$3.5568$ nb} \\ \hline
Interf.: 
 & $6.7\cdot 10^{-4}$ & $-3.58\cdot 10^{-3}$ & $-4.6\cdot 10^{-4}$ & 
   $-1.6\cdot 10^{-4}$ & $-1.6\cdot 10^{-4}$ & $-5.8\cdot 10^{-5}$ \\ \hline
$\sum$ interf.:    & \multicolumn{6}{c|}{$-3.74\cdot 10^{-3}$} \\ \hline
Rad.~tails: 
 & $1.145\cdot 10^{-2}$ & $9\cdot 10^{-5}$ & -- & -- & -- & -- \\ \hline
$\sum$ rad. tails: & \multicolumn{6}{c|}{$1.154\cdot 10^{-2}$} \\ \hline
Total: & \multicolumn{6}{c|}{$3.5845$ nb} \\ \hline\hline
 \multicolumn{7}{|l|}{$E_2 = 9.98$ GeV, $m_{\rm min}=9$ GeV:} \\ \hline
Continuum:  & \multicolumn{6}{c|}{$3.0666$ nb} \\ \hline
Interf.: 
 & $8.8\cdot 10^{-4}$ & $-4.1\cdot 10^{-3}$ & $-5.0\cdot 10^{-4}$ & 
   $-1.7\cdot 10^{-4}$ & $-1.6\cdot 10^{-4}$ & $-6.1\cdot 10^{-5}$ \\ \hline
$\sum$ interf.:    & \multicolumn{6}{c|}{$-4.14\cdot 10^{-3}$} \\ \hline
Rad.~tails: 
 & $1.328\cdot 10^{-2}$ & $1.04\cdot 10^{-4}$ & -- & -- & -- & -- \\ \hline
$\sum$ rad. tails: & \multicolumn{6}{c|}{$1.338\cdot 10^{-2}$} \\ \hline
Total: & \multicolumn{6}{c|}{$3.0949$ nb} \\ \hline
\end{tabular}
\end{center}
\label{table4}
\end{table}
\begin{figure}[htb]
\begin{center}
\leavevmode
\epsfxsize=12.cm
\epsffile[120 280 460 560]{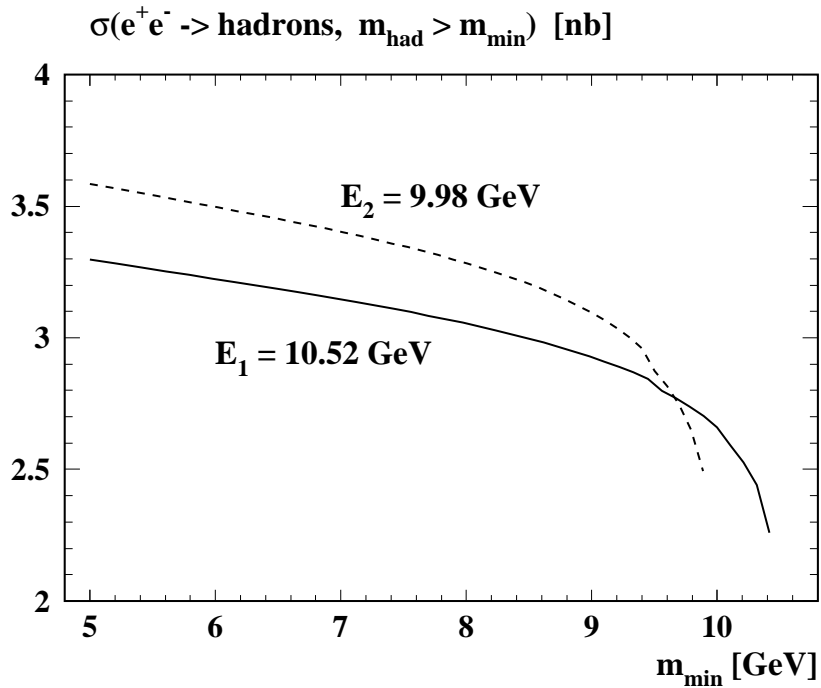}
\vskip -3mm
\caption[]{\label{fig4} {\em Dependence of the hadronic cross section 
$\sigma(e^+ e^- \to {\rm hadrons})$ on the cutoff in the minimal 
invariant mass of the hadronic system, $m_{\rm min}$, for the two 
energies $E_1 = 10.52$ GeV, $E_2 = 9.98$ GeV and $\alpha_s(M_Z) =
0.120$.}} 
\end{center}
\end{figure}
The interference 
between the $\Upsilon$ resonances and the virtual photon is completely
analogous to the leptonic case and leads to the same relative
corrections. However, the contributions from the radiative tails of
the Breit--Wigner amplitudes are distinctively different. Without 
initial state radiation the tail from one resonance is given by
\begin{equation}
\sigma_{\rm had, R} = 
\frac{12\,\pi\,\Gamma_e\,\Gamma_{\rm had}\,M^4}{s^3} 
\frac{M^2}{(s-M^2)^2+\Gamma^2 M^2}\,.
\label{eqtailhad}
\end{equation}
For the $\Upsilon(4,\,5,\,6)$ resonances an energy dependent width 
$\Gamma(s)$ has to be used in (\ref{eqtailhad}), which varies rapidly
in the region of interest. For $s$ below the $B{\overline B}$ threshold this
leads to a drastic suppression of these contributions. 
In the narrow width approximation eq.~(\ref{eqtailhad}) implies
\begin{equation}
\sigma_{\rm had, R} \Big|_{\rm NW} = 
\frac{12\,\pi^2\,\Gamma_e\,\Gamma_{\rm had}}{s} 
\frac{M}{\Gamma}\,\delta(s-M^2) \,. 
\label{eqtailhadnw}
\end{equation}
The contributions of interference terms and radiative tails from the
individual resonances to the hadronic cross section are listed in
Table~\ref{table4} for the energies $E_1 = 10.52$ GeV and $E_2 = 9.98$
GeV and $\alpha_s(M_Z^2) = 0.120$. Also given are the continuum
contribution and the sum. The cutoffs $m_{\rm min} = 5$~GeV and 9~GeV are
adopted. These correspond to
the minimal and maximal values recommended for a precision measurement
of $R_{\rm had}$. The relative strength of the interference terms is
identical for hadronic and leptonic final states. Again one observes
the accidental cancellation at $10.52$~GeV and the dominance of the
(negative) $\Upsilon(2)$ contribution at the energy $E_2 = 9.98$~GeV just
below this resonance. Compared to the leptonic cross section 
(Table~\ref{table3}) the radiative tails are significantly more 
prominent --- a consequence
of the relatively large direct, non--QED--mediated hadronic decay rates of
the $\Upsilon$ resonances. This leads to a positive correction of 2.3\%
and 1.2\% for $E_1$ and $E_2$, if $m_{\rm min} = 5$~GeV and to 2.7\%
and 1.3\% if $m_{\rm min} = 9$~GeV, respectively.
The difference in the strength of the $\Upsilon$
tails leads to an apparent variation of $R_{\rm had}$ around 1.3\%.
The cutoff dependence of the cross section is illustrated 
in Fig.~\ref{fig4}. 
The strong dependence of the cross section for tight cuts is again
clearly visible, suggesting a cut between about 5 and 9 GeV for $E_1$
and 5 and 8.5 GeV for $E_2$.
\vskip 1 cm
\noindent
{\Large 4. Uncertainties}\\

\noindent
From the comparison of different radiator functions for initial state
radiation the theoretical uncertainty from this source can be estimated
to be below one permille. The error induced through the present
simplified treatment of the hadronic vacuum polarization in the timelike
region is estimated around two permille and could easily be reduced even
further, if required. The combined theoretical uncertainty from these and
other effects is generously estimated below five permille. In addition
photons from initial state radiation might be included in the
invariant mass of the hadronic system or, conversely, photons from final
state radiation may escape the detection. This ${\cal O}(\alpha)$ effect
can only be evaluated for the concrete experimental analysis with the
help of a Monte Carlo simulation.
These theoretical uncertainties are significantly below the expected
experimental error of roughly two percent, which is dominated by 
systematical uncertainties \cite{DaveBessonpriv}.

\vskip 1 cm
\noindent
{\Large 5. Summary and Conclusions}\\

\noindent
Precise predictions have been presented for the total cross section in
the energy region explored presently by the CLEO experiment and at
a future $B$ meson factory. The present sample of nearly one million 
hadronic events allows for a small statistical error. These
measurements will determine the value for $\alpha_s$ under
particularly clean conditions similar to the $Z$ line shape
measurements but at a different energy. 
When compared to the experimental results from $Z$ decays, a
determination of $R$ with a precision of 2.5\% would evidently
demonstrate the running of $\alpha_s$ between 90 and 10 GeV. A precision
of 0.3\% would be competitive with the $\alpha_s$ measurements from the
$Z$ line shape which are based on the combined results of all four LEP
experiments.

\vskip 1 cm
\noindent
{\Large Acknowledgements}\\

\noindent
We thank Fred Jegerlehner and Helmut Burkhardt for providing their 
programs and Fred Jegerlehner for important discussions concerning 
the hadronic vacuum polarization. The interest of Dave Besson in this 
study was essential for its completion.
TT thanks the UK Particle Physics and Astronomy Research Council and the Royal
Society for support. 
This work was supported by BMFT under Contract 057KA92P(0), 
and INTAS under Contract INTAS-93-0744.
%
\def\app#1#2#3{{\it Act. Phys. Pol. }{\bf B #1} (#2) #3}
\def\apa#1#2#3{{\it Act. Phys. Austr.}{\bf #1} (#2) #3}
\def\lhc{Proc. LHC Workshop, CERN 90-10}
\def\npb#1#2#3{{\it Nucl. Phys. }{\bf B #1} (#2) #3}
\def\plb#1#2#3{{\it Phys. Lett. }{\bf B #1} (#2) #3}
\def\prd#1#2#3{{\it Phys. Rev. }{\bf D #1} (#2) #3}
\def\pR#1#2#3{{\it Phys. Rev. }{\bf #1} (#2) #3}
\def\prl#1#2#3{{\it Phys. Rev. Lett. }{\bf #1} (#2) #3}
\def\prc#1#2#3{{\it Phys. Reports }{\bf #1} (#2) #3}
\def\cpc#1#2#3{{\it Comp. Phys. Commun. }{\bf #1} (#2) #3}
\def\nim#1#2#3{{\it Nucl. Inst. Meth. }{\bf #1} (#2) #3}
\def\pr#1#2#3{{\it Phys. Reports }{\bf #1} (#2) #3}
\def\sovnp#1#2#3{{\it Sov. J. Nucl. Phys. }{\bf #1} (#2) #3}
\def\jl#1#2#3{{\it JETP Lett. }{\bf #1} (#2) #3}
\def\jet#1#2#3{{\it JETP Lett. }{\bf #1} (#2) #3}
\def\zpc#1#2#3{{\it Z. Phys. }{\bf C #1} (#2) #3}
\def\ptp#1#2#3{{\it Prog.~Theor.~Phys.~}{\bf #1} (#2) #3}
\def\nca#1#2#3{{\it Nouvo~Cim.~}{\bf #1A} (#2) #3}

%
\end{document}